# Tuning Coupled Toroidic and Polar Orders in a Bilayer Antiferromagnet

Chuangtang Wang[1,+], Xiaoyu Guo[1,+], Zixin Zhai[2], Meixin Cheng[3,4], Sang-Wook Cheong[5], Adam W. Tsen[3,4], Bing Lv[2], and Liuyan Zhao[1,*]

[1] Department of Physics, University of Michigan, Ann Arbor, USA
[2] Department of Physics, the University of Texas at Dallas, Richardson, TX, USA
[3] Department of Chemistry, University of Waterloo, Waterloo, ON, Canada
[4] Institute for Quantum Computing, University of Waterloo, Waterloo, ON, Canada
[5] Rutgers Center for Emergent Materials, Rutgers University, Piscataway, NJ, USA

[+] These authors contribute equally.
[*] Corresponding author email: lyzhao@umich.edu (L.Z.)

## Abstract

Magnetic toroidal order features a loop-like arrangement of magnetic dipole moments, thus breaking both spatial inversion ($P$) and time-reversal ($T$) symmetries while preserving their combined $PT$ symmetry. This $PT$ symmetry enables a linear magnetoelectric effect, allowing the coupling between magnetic toroidicity and electric polarity. However, the detection and control of two-dimensional (2D) magnetic toroidal order and the investigation of its linear magnetoelectric response remain largely unexplored. Here, using bilayer CrSBr as a platform, which hosts an in-plane layer-antiferromagnetic (AFM) order and simultaneously exhibits a magnetic toroidal order, we show compelling evidence for tuning this 2D magnetic toroidicity and its induced electric polarity through magnetic-field-dependent second harmonic generation (SHG). Under an out-of-plane magnetic field, we decompose the SHG signal into a time-reversal-odd component that scales with the magnetic toroidal moment and a time-reversal-even component that is proportional to the electric polarization. When sweeping the magnetic field from positive to negative values, we observe that the magnetic toroidicity retains its sign but diminishes in magnitude at higher fields while the electric polarity flips its sign and increases in strength at increasing fields below a critical threshold. When applying an in-plane electric field along the Néel vector direction, together with an out-of-plane field, we find that the magnetic toroidal and electric polar domains are moved in a locked fashion. These findings underscore the promise of 2D magnetic toroidal order in realizing giant linear magnetoelectric effects, opening exciting possibilities for next-generation electronic, magnetic, optical, and photonic devices enabled by 2D magnetoelectrics.

An electric dipole moment or an electric field breaks the spatial inversion (*P*) symmetry, while a magnetic dipole moment or a magnetic field violates the time-reversal (*T*) symmetry. In contrast, a magnetic toroidal moment breaks both *P* and *T* symmetries but preserves their product, *PT* symmetry[1–5]. Because of its symmetry property, it couples to mutually orthogonal electric and magnetic fields, or equivalently, to their cross product that respects *PT* symmetry[6–8]. Based on its orthogonal coupling fields, it supports off-diagonal linear magnetoelectric effect where the applied magnetic field and its induced electric polarization are orthogonal to each other[1,4]. To date, magnetic toroidal order and its associated linear magnetoelectric effect have been primarily studied in three-dimensional (3D) system for very few cases and remains pristine in the two-dimensional (2D) materials. However, with the rapid development of 2D magnetism, the realization of 2D magnetic toroidal order and the investigation of its magnetoelectric response are emerging as exciting frontiers[9,10].

Among the known 2D magnets, several *PT*-symmetric candidates have been identified, including layered AFMs such as even-layer $CrI_3$[11–13], even-layer CrSBr[14–18], and even-layer $MnBi_2Te_4$[19–23], as well as honeycomb AFMs such as $MnPS_3$[24,25] and $MnPSe_3$[24,26,27]. We choose the bilayer CrSBr as the material platform of this study as it hosts magnetic toroidal order[1,5]. It crystallizes in an orthorhombic lattice with the *mmm* point group. Below the Néel temperature ($T_N$) of 140 K[14], bilayer CrSBr enters a layered AFM phase with an *m'mm* magnetic point group, where spins align ferromagnetically along the *b*-axis within each layer and antiferromagnetically between the two layers (Fig. 1a). The order parameter of this layered AFM phase is a Néel vector along the *b*-axis ($\vec{N} = \vec{S}_1 - \vec{S}_2$), and equivalently, a magnetic toroidal moment along the *a*-axis ($\vec{T} = \hat{e}_{12} \times (\vec{S}_1 - \vec{S}_2) = \hat{e}_{12} \times \vec{N}$), where $\vec{S}_1$ and $\vec{S}_2$ are the spin moments in the first and second layers of bilayer CrSBr, and $\hat{e}_{12}$ is the unit vector connecting the two layers. (Fig. 1a). When magnetic fields are applied along different crystallographic directions, the layered AFM ground state evolves in distinct ways. For a field along the *a*-axis, the spins in both layers gradually reorient toward the field direction, forming a noncollinear spin configuration corresponding to the *22'2'* magnetic point group[28,29], as illustrated in Fig. 1b. This state eventually transitions to a fully polarized ferromagnetic (FM) state above a critical field of $B_a$ = 0.65 T at 80 K. When the field is along the *b*-axis, the layered AFM configuration remains stable until undergoing a spin-flip transition into a FM state with the *m'mm'* magnetic point group at $B_b$ = 0.13 T and 80 K, shown in Fig. 1c. For a magnetic field along the *c*-axis, the spins cant out of the plane, developing a different noncollinear texture associated with the *m'2'm* magnetic point group[30], as seen in Fig. 1d. Above a critical field of $B_c$ = 1.35 T at 80 K, this state transitions into a FM phase.

Among the spin states induced by applying magnetic fields along the three crystallographic axes, the noncollinear spin texture observed below $B_c$ is of particular interest for the study. This is because the time-reversal-invariant counterpart of its magnetic point group, *m'2'm*, corresponds to the polar point group *m2m*, which can support the emergence of an electric polarization ($\vec{P}$) along the *b*-axis in bilayer CrSBr, as indicated by the blue arrow in Fig. 1d. According to the Katsura–Nagaosa–Balatsky (KNB) mechanism[31], polarization arises via $\vec{P} \propto \hat{e}_{12} \times (\vec{S}_1 \times \vec{S}_2)$. Alternatively, this conclusion can be understood through Neumann's principle. For this *PT*-symmetric AFM ground state that supports a magnetic toroidal order in Fig. 1a and belongs to the *m'mm* magnetic point group, its linear magnetoelectric susceptibility tensor $\overleftrightarrow{\alpha}$ is off-diagonal, with only $\alpha_{bc}$ and $\alpha_{cb}$ being nonzero[32], allowing for the introduction of an electric polarization along the *b*-axis in response to a magnetic field applied along the *c*-axis. (see Supplementary Information Note S1). Furthermore, this linear magnetoelectric tensor directly relates to the magnetic toroidal moment via $\alpha_{ij} = \varepsilon_{ijk} T_k$, where $\overleftrightarrow{\varepsilon}$ is the Levi-Civita tensor, resulting in an antisymmetric $\overleftrightarrow{\alpha}$, i.e., $\alpha_{bc} = -\alpha_{cb}$[6]. Note that upon the application of the *c*-axis magnetic field, the magnetic toroidal moment, as well as the Néel vector, decreases until it vanishes above $B_c$. Thus, this *c*-axis magnetic field tunes both the magnetic toroidization and the electric polarization in distinct ways.

To capture the magnetic toroidal order and investigate its linear magnetoelectric effect in bilayer CrSBr, we employed rotation anisotropy second harmonic generation (RA SHG), a technique previously demonstrated to be sensitive to the layered AFM order in CrSBr[14] and widely recognized for its capability to detect electric polarization[33,34] (see Methods). We begin with the temperature dependence without a magnetic field. Figure 2a presents the temperature dependence of SHG intensity in bilayer CrSBr across the reported $T_N$ of 140 K[14], measured in the linear parallel channel at the peak polarization angle of the RA SHG pattern (indicated by red and blue arrows at $\varphi$ = 35° and 145° with respect to *b*-axis in Fig. 2b). Above $T_N$, the SHG signal is negligibly small due to the combination of inversion symmetry that suppresses the leading-order electric dipole (ED) contribution to SHG, and the out-of-plane two-fold ($C_{2c}$) rotational symmetry, which prohibits any SHG response under normal incidence. Below $T_N$, the SHG intensity increases because the layered AFM state with a finite toroidal moment (*m'mm*) breaks both the inversion and $C_{2c}$ symmetries to allow for ED SHG at normal incidence. The shape of the RA SHG pattern remains unchanged below $T_N$, aside from the intensity enhancement. A representative polar plot acquired at 80 K is shown in Fig. 2b, featuring four lobes of equal SHG intensity located symmetrically on the two sides of *a*- and *b*-axes. This RA SHG pattern is well fitted by the functional form of ED RA SHG under *m'mm*, where the ED SHG susceptibility tensor scales proportionally to the Néel vector ($\vec{N}$) and the magnetic toroidal moment ($\vec{T}$) (see Supplementary Information Note S2). Furthermore, the temperature dependence of the SHG intensity in the parallel polarization channel, $I_{\text{para}}^{2\omega}(T)$, shown in Fig. 2a, is fitted to the power-law form $I_{\text{para}}^{2\omega}(T) \propto |T - T_N|^{2\beta}$, yielding a Néel temperature $T_N$ = 141 $\pm$ 0.29 K, in good agreement with the reported value of 140 K[14], and a critical exponent $\beta$ = 0.26 $\pm$ 0.02, close to the 2D XY model value of 0.23[35,36].

We next examine the magnetic field dependence of SHG at 80 K, via applying magnetic fields along the three orthogonal crystallographic axes, namely, *a*-, *b*-, and *c*-axis. When the magnetic field (*B*) is applied along the *a*-axis (i.e., *B*//*a*), the SHG intensities of two representative lobes in the RA SHG pattern decrease uniformly and vanish above $B_a$ = 0.65 T (Fig. 2c). This behavior indicates that while the overall SHG intensity diminishes with increasing field, the pattern shape remains unchanged. A representative RA SHG polar plot taken at 0.4 T (*B* < $B_a$), shown in Fig. 2d, displays four symmetric lobes oriented identically to those at 0 T (Fig. 2b), albeit with reduced intensity. This result for *B*//*a* is consistent with a progressive canting of AFM spins away from the *b*-axis, fitted by ED RA SHG under the *22'2'* magnetic point group for *B* < $B_a$, eventually ending up as a fully polarized FM state along the *a*-axis, described by forbidden ED SHG under the centrosymmetric *mm'm'* point group for *B* > $B_a$ (see Supplementary Information Note S2). When *B*//*b*, both the SHG intensity and the RA SHG pattern remain constant until $B_b$ = 0.13 T, above which the SHG signal abruptly vanishes (Figs. 2e and 2f). This observation aligns with a spin-flip transition across $B_b$, where the system changes from an ED SHG-active layered AFM phase of *m'mm* magnetic point group to a centrosymmetric *m'mm'* FM phase that suppresses ED SHG (see Supplementary Information Note S2). In contrast, the magnetic field response for *B//c* is markedly different for these two former cases. The SHG intensity at $\varphi$ = 35° (red) shows a non-monotonic dependence on the increasing *B* field, while that at $\varphi$ = 145° (blue) decreases steadily, with both signals ultimately suppressed above $B_c$ = 1.35 T (Fig. 2g). These distinct trends indicate the presence of two pairs of uneven lobes in the RA SHG pattern and a substantial, continuous change of the pattern shape in the range 0 T < *B* < $B_c$. Notably, the RA SHG pattern at *B* < $B_c$, e.g., 0.8 T as shown in Fig. 2h, cannot be fitted solely using the ED SHG under the *m'2'm* magnetic point group (dashed lines), although the suppression of SHG above $B_c$ is consistent with the field-induced centrosymmetric FM state of the *m'm'm* point group.

To gain deeper insights into the RA SHG result under *B* < $B_c$ shown in Fig. 2g, we systematically tracked the evolution of RA SHG as the *B* field along the *c*-axis swept from positive (upwards) to negative (downwards) values. Fig. 3a presents the color map of the SHG intensity as functions of the *c*-axis *B* field from +1.5 T to -1.5 T and the polarization angle $\varphi$ from 0° to 360°, measured at 80K in the parallel channel ($I^{2\omega}_{\text{para}}(B,\varphi)$). A horizontal linecut at a fixed $\varphi$ reproduces the *B* dependence of SHG similar to those in Fig. 2g, while a vertical linecut at a specific *B* field yields the corresponding RA SHG pattern exemplified by the polar plot in Fig. 2h. Two key features emerge from this map. First, the four even lobes observed at 0 T gradually evolve into two pairs of asymmetric ones with the increasing *B* field, eventually disappearing beyond $B_c$ = ± 1.35 T. This trend aligns with the findings of Fig. 2g but offers a more complete visualization. Second, the asymmetry in the RA SHG lobes reverses with the sign of the *B* field. Under a positive field, the first and third lobes become stronger, while under a negative field, the second and fourth lobes are more prominent.

To quantitatively fit the RA SHG results at *B* < $B_c$, we focus on the two representative patterns at +0.8 T and -0.8 T shown in Fig. 3b. The analysis proceeds through the identification of SHG sources that are necessary and sufficient to explain the observed RA SHG features. First, the crystalline structure of bilayer CrSBr belongs to the centrosymmetric point group *mmm*, which forbids leading-order ED SHG contribution. Second, the layered AFM at *B* = 0 T, characterized by the *PT*-symmetric *m'mm* magnetic point group, activates ED SHG with a time variant (*c*-type) susceptibility tensor that scales linearly with the Néel vector $\vec{N}$ and the magnetic toroidal moment $\vec{T}$. Under a *c*-axis *B* field below $B_c$, the spins become noncollinear and lie within the *bc*-plane, $\vec{S}_i = S_{iy}\hat{b} + S_{iz}\hat{c}$ with *i* = 1 and 2 labeling the 1st and 2nd layer in the bilayer CrSBr. This spin configuration can be decomposed into two components, $\vec{M} = \vec{S}_1 + \vec{S}_2 = (S_{1z} + S_{2z})\hat{c}$, the net magnetization along the *c*-axis, and $\vec{N} = \vec{S}_1 - \vec{S}_2 = (S_{1y} - S_{2y})\hat{b}$ corresponding to $\vec{T} = \hat{e}_{12} \times \vec{N} = -(S_{1y} - S_{2y})\hat{a}$, the Néel vector along the *b*-axis together with the toroidal moment along the *a*-axis. The resulting magnetic point group becomes *m'2'm*, which permits ED SHG with a *c*-type susceptibility that continues to scale with $\vec{N}$ and $\vec{T}$ (i.e., $\chi^{\text{ED,(c)}}_{ijk} \propto \vec{N}$ and $\vec{T}$; see Supplementary Information Note S3). However, this alone cannot account for the RA SHG pattern asymmetry or its reversal under opposite *B* field directions. Third, to address the insufficiency of $\chi^{\text{ED,(c)}}_{ijk}$ in fitting the data, we extend the analysis to second order (quadratic) spin terms of the form $\vec{S}_1 \otimes \vec{S}_2$. For spins confined to the *bc*-plane, four independent combinations emerge. Among them, the antisymmetric combination $S_{1y}S_{2z} - S_{1z}S_{2y}$ transforms under the polar point group *m2m* and thus enables ED SHG through a time-invariant (*i*-type) susceptibility $\chi^{\text{ED,(i)}}_{ijk}$. Notably, this antisymmetric term directly corresponds to an electric polarization along the *b*-axis through the KNB mechanism, $\vec{P} \propto \hat{e}_{12} \times (\vec{S}_1 \times \vec{S}_2) = (S_{1y}S_{2z} - S_{1z}S_{2y})\hat{b}$, and accordingly, its induced *i*-type ED SHG scales with this polarization (i.e., $\chi_{ijk}^{\text{ED,(i)}} \propto \vec{P}$; see Supplementary Information Note S3). In contrast, the three symmetric quadratic terms, $S_{1y}S_{2z} + S_{1z}S_{2y}$, $S_{1y}S_{2y}$, and $S_{1z}S_{2z}$, all correspond to centro-symmetric point groups and therefore do not contribute to ED SHG. Taken together, the observed RA SHG patterns in Fig. 3b are successfully modeled as the coherent superposition between the *c*-type and *i*-type ED SHG contributions. We further comment that, as *B* reverses from +0.8 T to -0.8 T, the spin component along the *c*-axis flips sign, while the *b*-axis component remains unchanged. This leads to identical $\vec{N}$ and $\vec{T}$ (thus identical $\chi_{ijk}^{\text{ED,(c)}}$), but opposite $\vec{P}$ (thus opposite $\chi_{ijk}^{\text{ED,(i)}}$), as illustrated in the fit in Fig. 3b.

Having successfully identified and disentangled the two contributing sources to the RA SHG patterns in Fig. 3b, we now track their magnetic field evolution by applying the same fitting procedure across all fields presented in Fig. 3a. Figures 3c and 3d show the magnetic field dependence of $\chi_{yxy}^{\text{ED,(c)}}$ and $\chi_{yyy}^{\text{ED,(i)}}$, respectively, representative components of the *c*-type and *i*-type ED SHG susceptibility tensors (see Supplementary Information Note S4 for additional components). Below $B_c$, $\chi_{yxy}^{\text{ED,(c)}}$ is even in *B*, while $\chi_{yyy}^{\text{ED,(i)}}$ is odd, consistent with our earlier conclusion that $\vec{N}$ and $\vec{T}$ remain unchanged whereas $\vec{P}$ reverses direction when *B* is inverted along the *c*-axis. To model these field dependence, we constructed a fitting framework based on two key considerations: first, the out-of-plane net magnetization scales linearly with the *c*-axis magnetic field before $B_c$, implying that the angle $\theta$ between $\vec{S}_1$ and $\vec{S}_2$ satisfies $\cos\frac{\theta}{2} = \frac{B}{B_c}$; and second, to the first-order approximation, the *c*-type and *i*-type ED SHG susceptibility tensors scale linearly with $\vec{N}$(as well as $\vec{T}$) and $\vec{P}$, respectively. These lead to the following functional forms for the magnetic field dependence: $\chi_{yxy}^{\text{ED,(c)}} \propto \sin\frac{\theta}{2} = \sqrt{1 - \left(\frac{B}{B_c}\right)^2}$ and $\chi_{yyy}^{\text{ED,(i)}} \propto \sin\theta \propto \frac{B}{B_c}\sqrt{1 - \left(\frac{B}{B_c}\right)^2}$ (see Supplementary Information Note S5). Both expressions provide excellent fits to the experimental data in Figs. 3c and 3d. Notably, $\chi_{yyy}^{\text{ED,(i)}}$, and consequently $\vec{P}$, begins at zero at *B* = 0 T when $\theta$ = 180° for the layered AFM state, increases linearly with *B* at small fields, then reaches a maximum at *B* = 0.95 T when $\theta$ = 90° for orthogonally aligned $\vec{S}_1$ and $\vec{S}_2$, and finally vanishes at $B_c$ where $\theta$ = 0° in the fully polarized FM state. This behavior, illustrated in the inset of Fig. 3, captures the full progression of magnetic field-induced polarization in bilayer CrSBr.

After showing the magnetic field dependence of the magnetic toroidal and electric polar orders, we move forward to investigate the electric field control of them, by imaging their domain evolution upon the application of an in-plane electric field. Figure 4a top panel shows an optical image of a bilayer CrSBr electric device with the two electrodes separated along the *b*-axis, and bottom panel illustrates the experimental geometry with a *b*-axis electric field and a *c*-axis magnetic field of +0.8 T. Figure 4b displays a scanning SHG micrograph for a bilayer CrSBr taken at *B* = 0.8T, and *T* = 80 K, for $\varphi = 35°$ in the parallel channel for the bilayer region outlined by the white dashed box in Fig. 4a. The key feature is a dark line with a suppressed SHG intensity, highlighted by the black arrow, that separates the image into two regions labeled as Domain 1 and Domain 2. They correspond to two AFM states with opposite $\vec{N}$, and thereby, magnetic toroidization with opposite $\vec{T}$ and induced electric polarization with opposite $\vec{P}$ under an out-of-plane *B* field. The suppression in SHG intensity originates from the destructive interference between the SHG fields from the two domains that have a 180° optical phase difference[2,37,38]. We can digest this better by seeing the RA SHG pattern from the two domains as shown in Fig. 4c. While their intensity polar plots look nearly identical, their *c*-type and *i*-type SHG susceptibility tensors have opposite signs determined by the opposite directions of $\vec{N}$ ($\vec{T}$) and $\vec{P}$ between the two domain states, and therefore, their SHG optical phase differ by 180° at all $\varphi$. These domains randomly and spontaneously populate through thermal cycles with a typical size of several $\mu$m. With the +0.8T magnetic field, after applying a voltage drop of 4V across the electrodes for 60s, we notice that the toroidal and polar domains remain unchanged, as shown in Fig. 4d. But by elevating the voltage up to 20V and keeping it for 60s, we can clearly see the change of the domain distribution shown in Fig. 4e, resulting in one dominant domain with $\vec{P}$ pointing the +*b* direction across the entire region. This can be understood equivalently by either the electric field coupling to the electric polar order or the cross product of electric and magnetic field controlling the magnetic toroidal order. Furthermore, by flipping the voltage drop to the -*b* direction while maintaining the magnetic field direction for another 60s, we then observe a nearly even population of the two types of domains shown in Fig. 4f.

In summary, we successfully captured the magnetic toroidal order and further demonstrated its linear magnetoelectric effect in this *PT*-symmetric 2D magnet, bilayer CrSBr, by employing ED SHG to disentangle the contributions from the magnetic toroidal moment and the induced electric polarization, to track their evolution under an applied magnetic field, and furthermore, to demonstrate the tunability by the external electric and magnetic fields. While our experiment did not directly quantify the magnetoelectric coupling constant, theoretical predictions suggest it is significant[10], consistent with our observation of an *i*-type contribution from $\vec{P}$ being comparable in magnitude to the *c*-type contribution from $\vec{N}$ ($\vec{T}$) in the ED SHG response. Moreover, considering the robust magnetism[14,39–41] and the remarkable electronic[42,43], excitonic[44–46], and polaritonic[47–50] properties of bilayer CrSBr, this linear magnetoelectric effect could serve as a key mechanism for coupling the spin and charge/lattice degrees of freedom, which may, in turn, enrich its electric, optical, and photonic responses under magnetic fields applied along the *c*-axis. In particular, the condition for the maximum induced $\vec{P}$, *i.e.*, orthogonally aligned spins in the two layers under the *c*-axis magnetic field, is the same as that for the largest exciton nonlinearity[51] and the strongest magnon oscillations[52] in bilayer CrSBr. Finally, we note that integrating bilayer CrSBr into functional devices exploiting this magnetic toroidal order-induced linear magnetoelectric effect is both timely and promising.

**Methods**

**Crystal growth and sample fabrication**

CrSBr single crystals were grown by a direct solid–vapor technique in a box furnace in sealed quartz containers using Cr power (Alfa Aesar, 99.97%), S powder (Alfa Aesar, 99.5%) and solidified $Br_2$ (Alfa Aesar 99.8%) using liquid nitrogen[39]. The Cr:S:Br molar ratio is 1 : 1.1 : 1.2 where the slight excess of S and $Br_2$ creates a positive vapor pressure that both minimizes defect formation and promotes larger crystal growth. The sealed ampoule was heated slowly to 930 °C, held for 20 h, then cooled at 1 °C/h down to 750 °C before being quenched to room temperature. Large CrSBr crystals form naturally at the bottom of the quartz ampoule.

Bilayer CrSBr samples were mechanically exfoliated with scotch tapes from the bulk single crystals onto 90 nm $SiO_2$/Si substrates inside a nitrogen-filled glovebox with water and oxygen levels below 0.1 ppm. To avoid the nonlinear optical signal from hBN, our bilayer CrSBr samples were not encapsulated by hBN flakes but were transferred into the optical cryostat (AttoDry 2100) with a short 2-minute exposure to the ambient environment. The thickness of the bilayer CrSBr was determined by optical contrast and verified by atomic force microscopy. The Au electrodes were pre-fabricated on $SiO_2$/Si substrates for electric field control devices. Bilayer CrSBr was picked up by polymer stamp and transferred onto the pre-patterned electrodes inside the glovebox.

**Rotation anisotropy second harmonic generation (RA SHG)**

RA SHG measurements were conducted under a normal incidence geometry with a femtosecond laser with 80 MHz repetition rate and 100 fs pulse duration. Linearly polarized fundamental light (800 nm wavelength) was focused onto the bilayer CrSBr sample as a ~1 $\mu$m spot. The reflected SHG light (400 nm in wavelength) was collected and directed to the gated photon counting PMT with an analyzer in the front. The polarizations of the fundamental and the SHG light was set to be either parallel (parallel channel) or orthogonal (cross channel) to each other and the polarization angle was rotated to construct the RA SHG polar plot. All field-dependent RA-SHG measurements were carried out at 80 K inside the attoDry 2100 cryostat. The magnetic field dependent measurements along *a*- and *b*-axes were performed in the Voigt geometry while measurements along *c*-axis were done in the Faraday geometry. The SHG scanning measurements were carried out inside attoDry 2100 with a xyz piezo stage.

**Data availability**

All data supporting this work are available from the corresponding author upon request.

**Acknowledgements**

L.Z. acknowledges support from the U.S. Department of Energy (DOE), Office of Science, Basic Energy Science (BES), under award No. DE-SC0024145 (for bilayer sample fabrication, temperature, magnetic field, and electric field dependent SHG measurements). B.L. and Z. Z. acknowledges support from US Air Force Office of Scientific Research Grant No. FA9550-19-1-0037, National Science Foundation- DMREF-2324033, and Office of Naval Research grant no. N00014-23-1-2020 (for bulk single crystal growth). A.W.T. acknowledges support from the Dorothy Killam Fellowship and US Air Force Office of Scientific Research Grant No. FA9550-24-1-0360 (for bilayer electric device fabrication). S.W.C. acknowledges support from the W. M. Keck foundation grant to the Keck Center for Quantum Magnetism at Rutgers University.

**Author contribution**

C.W., X.G., and L.Z. conceived and initiated the project. C.W. fabricated the bilayer CrSBr samples, and performed the temperature-, magnetic field-, and electric field-dependent SHG measurements. M.C. fabricated the bilayer CrSBr devices for the electric field control under the guidance of A.W.T.. Z.Z. and B.L. provided the bulk CrSBr single crystals. C.W., X.G., S.W.C., and L.Z. analyzed the experimental data and prepared the manuscript. All authors participated in discussions of the results.

**Competing interest**

The authors declare no competing interests.

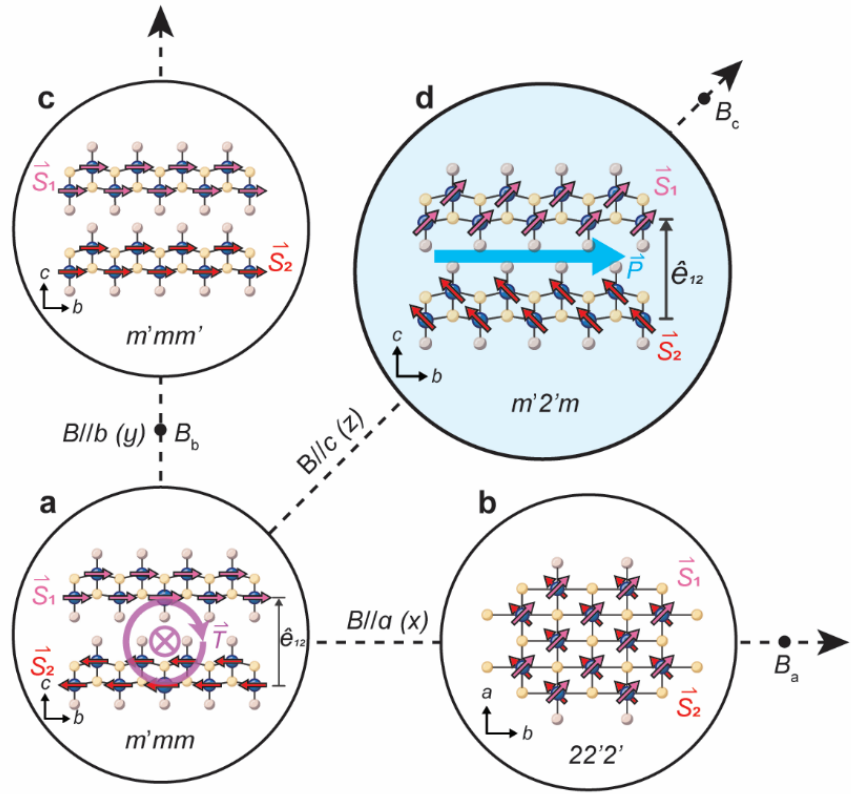

**Fig.1 | Schematic of the evolution of spin configuration under applied magnetic fields. a,** The ground state of bilayer CrSBr, with spins in layer one, $\vec{S_1}$ (pink), and layer two, $\vec{S_2}$ (red), aligned parallel along the *b*-axis within each layer and antiparallel between the two layers, with a magnetic point group *m'mm*, supporting magnetic toroidal moment $\vec{T}$ (with toroidal circulation direction indicated as purple arrow) along *a*-axis. **b,** For $B < B_a$, the spins are canted in the *ac*-plane towards the *a*-axis, with a magnetic point group *22'2'*. **c,** For $B < B_b$, the spins remain the same as B = 0 T; and for $B > B_b$, the spins flip and lead to a FM state with a magnetic point group *m'mm'*. **d,** For $B < B_c$, the noncollinear spins belong to a reduced magnetic point group *m'2'm* that supports the generation of an electric polarization (blue arrow) along the *b*-axis based on the Katsura–Nagaosa–Balatsky mechanism, $\vec{P} \propto \hat{e}_{12} \times (\vec{S_1} \times \vec{S_2})$. with $\hat{e}_{12}$ being the unit separation vector between the two layers.

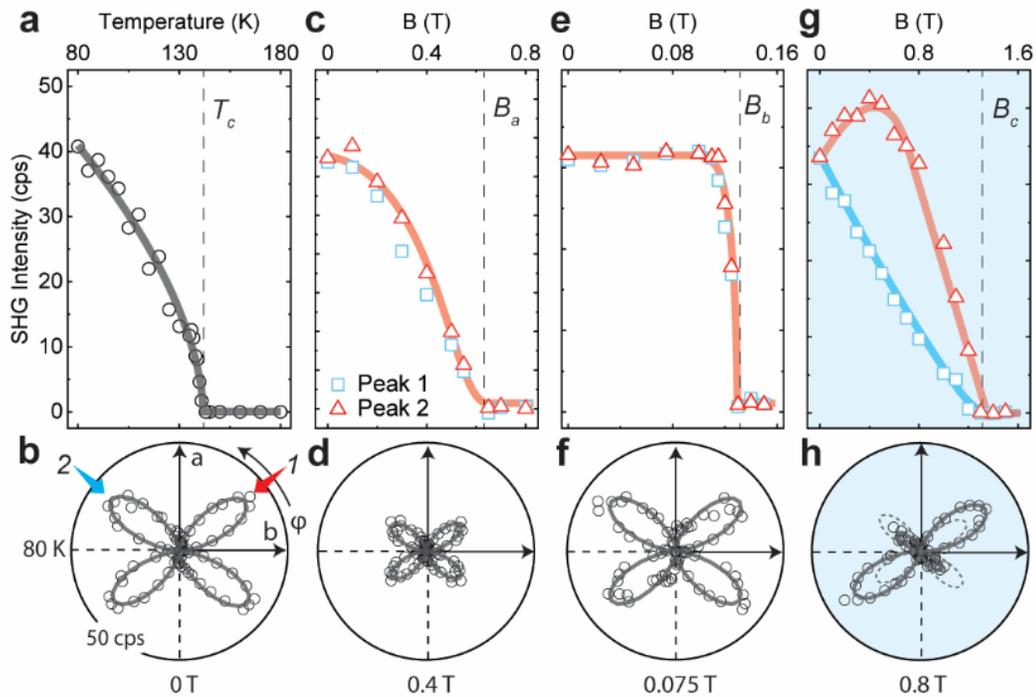

**Fig.2 | Temperature and magnetic field dependence of SHG. a,** Temperature dependent SHG intensity in the parallel channel. The solid line is a fit to the function form for a second order phase transition with a fitted transition temperature $T_N$ =141 K $\pm$ 0.29. **b,** RA SHG pattern at 80 K and 0 T. Red and blue arrows mark the RA SHG peak angles of 35° and 145° with respect to *b*-axis, respectively. Magnetic-field dependent SHG intensities in the parallel channel at 35° and 145° at 80 K are shown in **c**, with *B* along the *a*-axis, **e**, with *B* along the *b*-axis, and **g**, with *B* along the *c*-axis, and their corresponding representative RA SHG patterns at selected magnetic fields shown in **d, f, h**, respectively. The dashed plot in **h** is the fitting with only *c*-type ED SHG contribution. All polar plots use the same intensity scale 50 cps.

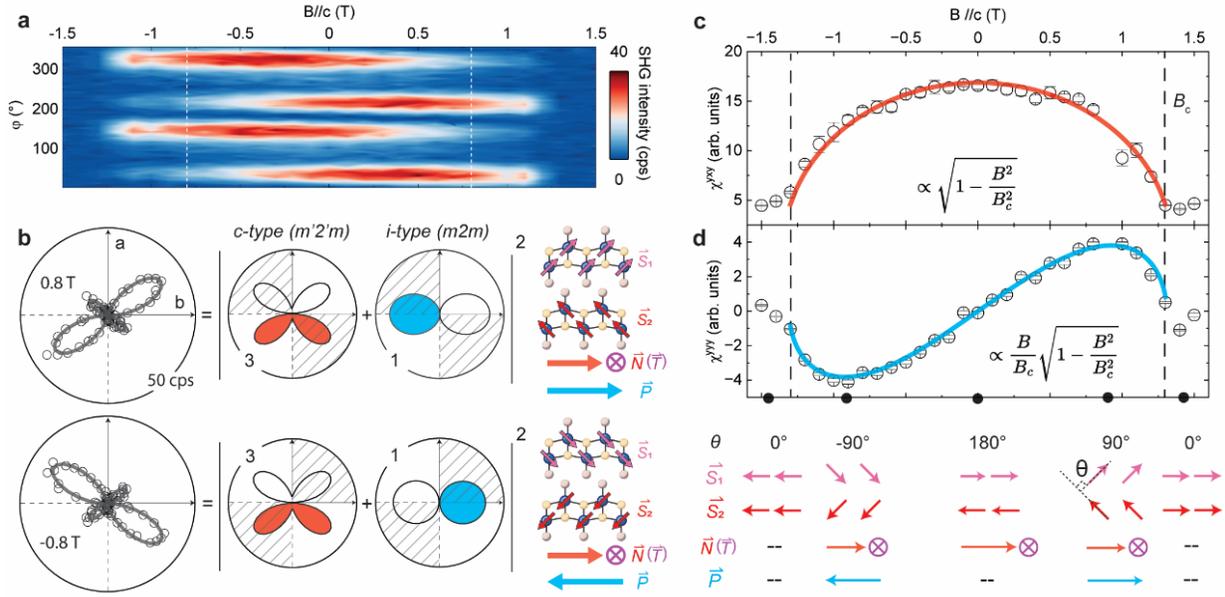

**Fig.3 | Identification and separation of *c*-type and *i*-type RA ED SHG under the *c*-axis magnetic field. a**, Contour plot of the SHG intensity taken in the parallel channel as functions of an magnetic field along the *c*-axis (horizontal axis) and the incident polarization angle (vertical axis). The dash white lines mark +0.8 T and -0.8 T. **b**, RA SHG patterns at +0.8 T and -0.8 T fitted with a coherent superpositon between the *c*-type ($m'2'm$) and *i*-type ($m2m$) ED SHG contributions. The colored shaded area indicates an optical phase shift of π relateive to the white areas. The stripped shaded area indicates the destructive interference to produce a suppressed SHG intensity. The inset shows the spin configuration illustrations at +/-0.8 T, with their Néel vectors (magnetic toroidal moments) aligned but electric polarizations flipped directions. **c**, Extracted $\chi_{yxy}^{ED,(c)}$ as a function of the magnetic field B along the c-axis. The solid line is the fit to the function form $\propto \sqrt{1 - \frac{B^2}{B_c^2}}$. **d**, Fitted $\chi_{yyy}^{ED,(i)}$ as a function of B. The solid line is the fit to the function form $\propto \frac{B}{B_c}\sqrt{1 - \frac{B^2}{B_c^2}}$. Error bars represent standard errors computed from 500 boostrap resamples of the RA SHG data. The insets illustrate the spin configurations, their Néel vectors (magnetic toroidal moments) and their corresponding induced electric polarization at selective magnetic fields of +1.5 T, +0.95 T, 0 T, -0.95 T, and -1.5 T.

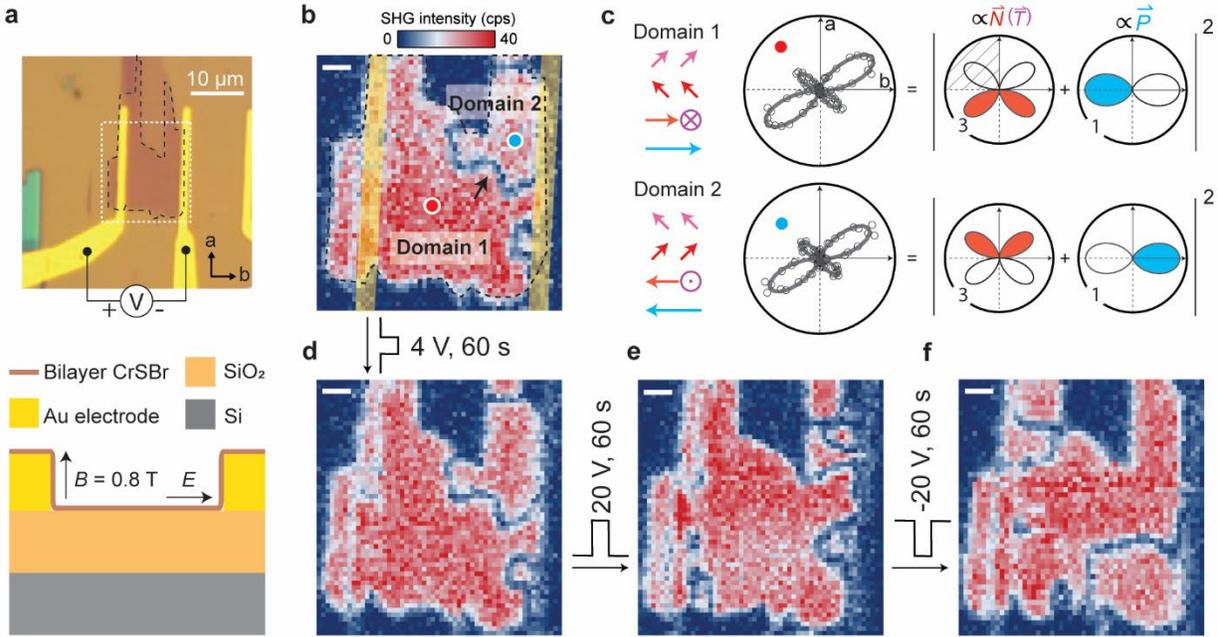

**Fig.4 | Electric field control of the toroidic and polar domains. a,** The optical image (top) and the schematic illustration (bottom) showing the experimental geometry for the electric field control of magnetic toroidal and electric polar domains. A bilayer CrSBr is transferred onto pre-patterned Au electrodes, across which voltage pulses are applied along the *b*-axis. A magnetic field of +0.8 T is applied along the *c*-axis. The dashed black line outlines the bilayer CrSBr and the white dashed line marks the region where scanning SHG measurement shown in **b** is performed. **b,** Scanning SHG micrograph of a bilayer CrSBr taken at $\varphi = 35°$ in the linear parallel channel, with $B = +0.8$ T applied along *c*-axis. It shows two domains that are separated by a line of suppressed SHG intensity as arrow indicated. The bilayer region is outlined by the black dashed line. The yellow region indicates the Au electrode. The red and blue dots indicate the position where the RA SHG patterns in **c** were taken. Scale bar, 2 μm. **c,** RA SHG polar plots measured from the two domains and fitted with a coherent superpositon between the *c*-type (scaled with Néel vector or magnetic toroidal moment) and *i*-type (scaled with electric polarization) ED SHG contributions. The insets show the illustration of the noncollinear spins in the two domains, as well as their associated Néel vectors (magnetic toroidal moment) and induced electric polarizations, both of which are opposite between the two domains. **d, e, f,** Consecutive domain images after appling 60 seconds voltage pulse of 4 V, 20 V, and -20 V, respectively. No domain motion is observed under 4V, whereas domains move signicantly under 20 V and -20 V pulses.

Supplementary Information for

# Tuning Coupled Toroidic and Polar Orders in a Bilayer Antiferromagnet

Chuangtang Wang[1,+], Xiaoyu Guo[1,+], Zixin Zhai[2], Meixin Cheng[3,4], Sang-Wook Cheong[5], Adam W. Tsen[3,4], Bing Lv[2], and Liuyan Zhao[1,*]

[1] Department of Physics, University of Michigan, Ann Arbor, USA
[2] Department of Physics, the University of Texas at Dallas, Richardson, TX, USA
[3] Department of Chemistry, University of Waterloo, Waterloo, ON, Canada
[4] Institute for Quantum Computing, University of Waterloo, Waterloo, ON, Canada
[5] Rutgers Center for Emergent Materials, Rutgers University, Piscataway, NJ, USA

[+] These authors contribute equally.
[*] Corresponding author email: lyzhao@umich.edu (L.Z.)

**Note 1: Magnetoelectric susceptibility tensor**

Starting from the constitutive relation: $P_i = \vec{\alpha}_{ij} H_j$, where $P_i$ is the polarization (polar), $H_j$ is the magnetic field (axial), and $\vec{\alpha}_{ij}$ is an axial tensor. Under a spatial symmetry operation $R$, $\vec{\alpha}_{ij}$ must satisfy Neumann's principle:

$$\vec{\alpha}_{ij} = (detR) R_{ik} R_{jl} \vec{\alpha}_{kl}$$

While under a spatial symmetry combined time-reversal symmetry $T$ (primed), $R' = TR$,

$$\vec{\alpha}_{ij} = -(detR) R_{ik} R_{jl} \vec{\alpha}_{kl}$$

to account for

$$H \xrightarrow{T} -H, P \xrightarrow{T} P.$$

Therefore, for a magnetic point group of *m'mm*, we enforce invariance under each of three symmetry operations: $m'_x$, $m_y$, $m_z$, with $R_x = diag(-1,1,1)$, $R_y = diag(1,-1,1)$ and $R_z = diag(1,1,-1)$, respectively. Thus, the only nonzero components are $\alpha_{yz}$ and $\alpha_{zy}$.

Therefore, its magnetoelectric susceptibility tensor $\vec{\alpha}$ has the form of:

$$\vec{\alpha} = \begin{pmatrix} 0 & 0 & 0 \\ 0 & 0 & \alpha_{bc} \\ 0 & \alpha_{cb} & 0 \end{pmatrix}.$$

**Note 2: Functional forms for ED SHG for point group $m'mm$, $22'2'$, $m'2'm$ and $m2m$**

The *c*-type ED SHG nonlinear optical susceptibility tensors for $m'mm$, $22'2'$, and $m'2'm$ are the same. They are polar and of rank-3:

$$\chi_{m'mm/22'2'/m'2'm}^{ED,(c)}(2\omega) = \begin{pmatrix} \begin{pmatrix} \chi^{xxx} \\ 0 \\ 0 \end{pmatrix} & \begin{pmatrix} 0 \\ \chi^{xyy} \\ 0 \end{pmatrix} & \begin{pmatrix} 0 \\ 0 \\ \chi^{xzz} \end{pmatrix} \\ \begin{pmatrix} 0 \\ \chi^{yxy} \\ 0 \end{pmatrix} & \begin{pmatrix} \chi^{yxy} \\ 0 \\ 0 \end{pmatrix} & \begin{pmatrix} 0 \\ 0 \\ 0 \end{pmatrix} \\ \begin{pmatrix} 0 \\ 0 \\ \chi^{zxz} \end{pmatrix} & \begin{pmatrix} 0 \\ 0 \\ 0 \end{pmatrix} & \begin{pmatrix} \chi^{zxz} \\ 0 \\ 0 \end{pmatrix} \end{pmatrix}$$

In case of normal incidence, the resulted functional forms for RA-SHG in the parallel and the crossed channels are:

$$I_{parallel}^{ED,(c)}(2\omega) \propto ((\chi^{xyy} + 2\chi^{yxy})cos(\phi)^2 sin(\phi) + \chi^{xxx} sin(\phi)^3)^2$$

$$I_{crossed}^{ED,(c)}(2\omega) \propto (\chi^{xyy} cos(\phi)^3 + (\chi^{xxx} - 2\chi^{yxy})cos(\phi)sin(\phi)^2)^2$$

The $m2m$ supports *i*-type ED SHG radiation with the corresponding susceptibility tensor,

$$\chi_{m2m}^{ED,(i)}(2\omega) = \begin{pmatrix} \begin{pmatrix} 0 \\ \chi^{xxy} \\ 0 \end{pmatrix} & \begin{pmatrix} \chi^{xxy} \\ 0 \\ 0 \end{pmatrix} & \begin{pmatrix} 0 \\ 0 \\ 0 \end{pmatrix} \\ \begin{pmatrix} \chi^{yxx} \\ 0 \\ 0 \end{pmatrix} & \begin{pmatrix} 0 \\ \chi^{yyy} \\ 0 \end{pmatrix} & \begin{pmatrix} 0 \\ 0 \\ \chi^{yzz} \end{pmatrix} \\ \begin{pmatrix} 0 \\ 0 \\ 0 \end{pmatrix} & \begin{pmatrix} 0 \\ 0 \\ \chi^{zyz} \end{pmatrix} & \begin{pmatrix} 0 \\ \chi^{zyz} \\ 0 \end{pmatrix} \end{pmatrix}$$

Under normal incidence, the resulting functional forms for RA-SHG in the parallel and the crossed channels are:

$$I_{parallel}^{ED,(i)}(2\omega) \propto (\chi^{yyy} cos(\phi)^3 + (\chi^{yxx} + 2\chi^{xxy})cos(\phi)sin(\phi)^2)^2$$

$$I_{crossed}^{ED,(i)}(2\omega) \propto ((2\chi^{xxy} - 2\chi^{yyy})cos(\phi)^2 sin(\phi) - \chi^{yxx} sin(\phi)^3)^2$$

Note that the *c*-type and the *i*-type have the complementary nonzero tensor elements, corresponding to orthogonal polarization components rotated by 90°. Therefore, the interference between *c*-type and *i*-type contributions is straightforward, as their polarization components add vectorially in orthogonal bases:

$$I_{parallel}^{ED,interfered}(2\omega) \propto ((\chi^{xyy} + 2\chi^{yxy})\cos(\phi)^2\sin(\phi) + \chi^{xxx}\sin(\phi)^3$$
$$+\chi^{yyy}\cos(\phi)^3 + (\chi^{yxx} + 2\chi^{xxy})\cos(\phi)\sin(\phi)^2)^2$$
$$I_{crossed}^{ED,interfered}(2\omega) \propto (\chi^{xyy}\cos(\phi)^3 + (\chi^{xxx} - 2\chi^{yxy})\cos(\phi)\sin(\phi)^2$$
$$+(2\chi^{xxy} - 2\chi^{yyy})\cos(\phi)^2\sin(\phi) - \chi^{yxx}\sin(\phi)^3)^2$$

**Note 3: Group theory analysis of spin terms**

Consider the spins in the two layers: $\vec{S_1} = (S_{1x}, S_{1y}, S_{1z})$ and $\vec{S_2} = (S_{2x}, S_{2y}, S_{2z})$, where the number in the subscript denotes the top layer (1) and bottom layer (2), and the letter in the subscript denotes the corresponding component in the Cartesian coordinate. The linear combination of the linear spin terms and their corresponding irreducible representation (irrep) of the $mmm$ are tabulated in Table S1. The table is built by constructing linear combinations of the spin terms and examine their transformations under each symmetry operations in the point group $mmm$ to fill in the characters. By comparing the filled characters with the character table, we are able to determine the irrep, the basis and the corresponding magnetic point group of each linear combination. Each linear combination of the linear terms corresponds to a spin configuration shown in the table.

**Table S1.** *Character table for linear spin combinations*

| Linear spin | Irrep | Basis | Subgroup | $m_x$ | $m_y$ | $m_z$ | $2_x$ | $2_y$ | $2_z$ | $i$ | Spin config |
|---|---|---|---|---|---|---|---|---|---|---|---|
| $S_{1x} + S_{2x}$ | $B_{3g}$ | $R_x$ | $mm'm'$ | 1 | -1 | -1 | 1 | -1 | -1 | 1 | FM along a |
| $S_{1y} + S_{2y}$ | $B_{2g}$ | $R_x$ | $m'mm'$ | -1 | 1 | -1 | -1 | 1 | -1 | 1 | FM along b |
| $S_{1z} + S_{2z}$ | $B_{1g}$ | $R_x$ | $m'm'm$ | -1 | -1 | 1 | -1 | -1 | 1 | 1 | FM along c |
| $S_{1x} - S_{2x}$ | $B_{2u}$ | y | $mm'm$ | 1 | -1 | 1 | -1 | 1 | -1 | -1 | AFM along a |
| $S_{1y} - S_{2y}$ | $B_{3u}$ | x | $m'mm$ | -1 | 1 | 1 | 1 | -1 | -1 | -1 | AFM along b |
| $S_{1z} - S_{2z}$ | $A_u$ | xyz | $m'm'm'$ | -1 | -1 | -1 | 1 | 1 | 1 | -1 | AFM along c |

Next, to identify all the possible bilinear terms, we considered the out productor of the two spins,

$$S_1 \otimes S_2 = \begin{pmatrix} S_{1x}S_{2x} & S_{1x}S_{2y} & S_{1x}S_{2z} \\ S_{1y}S_{2x} & S_{1y}S_{2y} & S_{1y}S_{2z} \\ S_{1z}S_{2x} & S_{1z}S_{2y} & S_{1z}S_{2z} \end{pmatrix}$$

By projecting the linear combinations of the bilinear terms into the irrep of $mmm$, we constructed Table S2.

**Table S2.** *Character table for bilinear spin combinations*

| Linear spin | Irrep | Basis | Subgroup | $m_x$ | $m_y$ | $m_z$ | $2_x$ | $2_y$ | $2_z$ | $i$ |
|---|---|---|---|---|---|---|---|---|---|---|
| $S_{1y}S_{2z} + S_{1z}S_{2y}$ | $B_{3g}$ | $R_x$ | $2/m$ | 1 | -1 | -1 | 1 | -1 | -1 | 1 |
| $S_{1x}S_{2z} + S_{1z}S_{2x}$ | $B_{2g}$ | $R_x$ | $2/m$ | -1 | 1 | -1 | -1 | 1 | -1 | 1 |
| $S_{1x}S_{2y} + S_{1y}S_{2x}$ | $B_{1g}$ | $R_x$ | $2/m$ | -1 | -1 | 1 | -1 | -1 | 1 | 1 |
| $S_{1y}S_{2z} - S_{1z}S_{2y}$ | $B_{2u}$ | y | $m2m$ | 1 | -1 | 1 | -1 | 1 | -1 | -1 |
| $S_{1x}S_{2z} - S_{1z}S_{2x}$ | $B_{3u}$ | x | $2mm$ | -1 | 1 | 1 | 1 | -1 | -1 | -1 |
| $S_{1x}S_{2y} - S_{1y}S_{2x}$ | $A_u$ | xyz | $222$ | -1 | -1 | -1 | 1 | 1 | 1 | -1 |
| $S_{1x}S_{2x}, S_{1y}S_{2y}, S_{1z}S_{2z}$ | $E$ | $x^2, y^2, z^2$ | $mmm$ | 1 | 1 | 1 | 1 | 1 | 1 | 1 |

**Note 4: Magnetic field dependence of all tensor elements for B//c case**

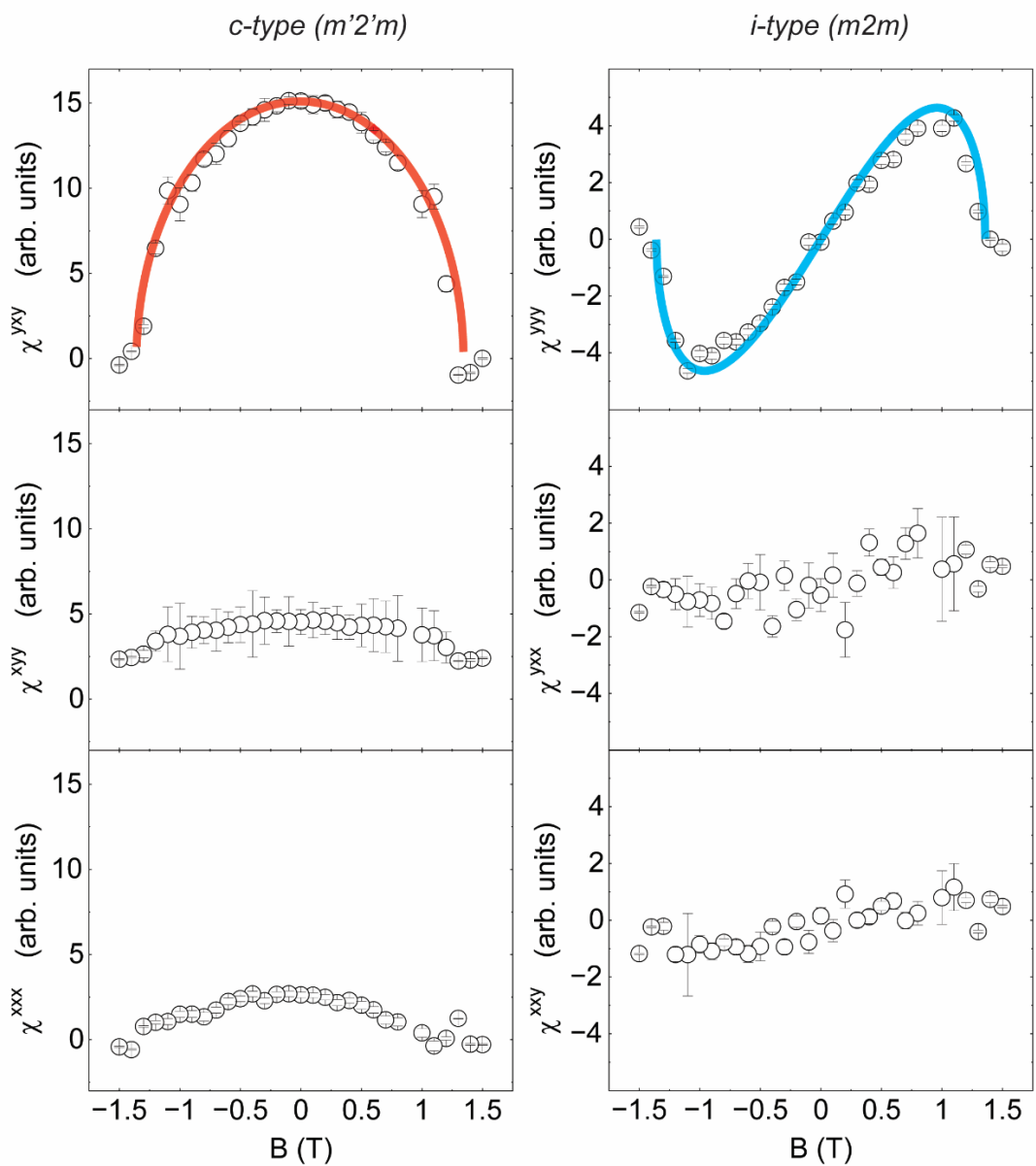

**Fig. S1** Fitted tensor elements vs magnetic field. The left column are three tensor elements corresponding to the *c*-type SHG contribution and the right column are three tensor elements corresponding to the *i*-type SHG contribution.

**Note 5: Fitting model for field-dependence of nonlinear susceptibility tensor elements**

To comprehend these two trends, we build the following fitting model. By defining the angle between $\vec{S}_1$ and $\vec{S}_2$ to be $\theta$, we have $\vec{M} = 2S \cos\frac{\theta}{2}\hat{c}$, $\vec{N} = 2S \sin\frac{\theta}{2}\hat{b}$, and $\vec{P} \propto S^2 \sin\theta$, where $S$ represents the magnetic moment magnitude. Knowing that the out-of-plane net magnetization scale linearly with the c-axis $B$ field, i.e., $M = \gamma B$, and that the fully polarized FM magnetic state is achieved at $B_c$ with a net magnetization of $2S$, i.e., $2S = \gamma B_c$, we arrive at $\cos\frac{\theta}{2} = \frac{B}{B_c}$, and thereby, $\vec{N} = 2S\sqrt{1 - \left(\frac{B}{B_c}\right)^2}$ and $\vec{P} \propto \cos\left(\frac{\theta}{2}\right)\sin\left(\frac{\theta}{2}\right) = \frac{B}{B_c}\sqrt{1 - \left(\frac{B}{B_c}\right)^2}$